\documentclass[twocolumn]{aastex631}

\newcommand{\bmath}[1]{\mbox{\boldmath{$#1$}}}

\definecolor{darkbrown}{HTML}{8c4600}
\definecolor{darkblue}{HTML}{1833a1}
\newcommand{\m}{{\rm m}}
\newcommand{\s}{{\rm s}}

\newcommand{\treinf}{t_{\rm reinf}}
\newcommand{\tstop}{t_{\rm stop}}

\newcommand{\vk}{v_{\rm k}}

\newcommand{\Omegainv}{\Omega^{-1}}
\newcommand{\Sigmag}{\Sigma_{\rm g}}
\newcommand{\Sigmap}{\Sigma_{\rm p}}

\newcommand{\cs}{c_{\rm s}}
\newcommand{\rhos}{\rho_{\rm s}}
\newcommand{\rhop}{\rho_{\rm p}}
\newcommand{\rhog}{\rho_{\rm g}}

\submitjournal{ApJ}

\graphicspath{{./}{plots/}}

\shorttitle{}
\shortauthors{Carrera and Simon}
\begin{document}

%% In v6.31 you can include a footnote in the title.

\title{The Streaming Instability Cannot Form Planetesimals from mm-size Grains in Pressure Bumps}

%% A significant change from earlier AASTEX versions is in the structure for 
%% calling author and affiliations. The change was necessary to implement 
%% auto-indexing of affiliations which prior was a manual process that could 
%% easily be tedious in large author manuscripts.
%%
%% The \author command is the same as before except it now takes an optional
%% argument which is the 16 digit ORCID. The syntax is:
%% \author[xxxx-xxxx-xxxx-xxxx]{Author Name}
%%
%% This will hyperlink the author name to the author's ORCID page. LaTeX will
%% do some limited checking of the format of the ID to make sure it is valid.
%% If the "orcid-ID.png" image file is present or in the LaTeX pathway, the
%% OrcID icon will appear next to the authors name.
%%
%% Use \affiliation for affiliation information.
\author[0000-0001-6259-3575]{Daniel Carrera}
\affiliation{Department of Physics and Astronomy, Iowa State University, Ames, IA, 50010, USA}

\author[0000-0002-3771-8054]{Jacob B. Simon}
\affiliation{Department of Physics and Astronomy, Iowa State University, Ames, IA, 50010, USA}

%% The new \altaffiliation can be used to indicate some secondary information
%% such as fellowships. This command produces a non-numeric footnote that is
%% set away from the numeric \affiliation footnotes.  It must come BEFORE the
%% \affiliation call, right after the \author command, in order to place the
%% footnotes in the proper location.

%% It is the author's responsibility to make sure the \correspondingauthor
%% is also in the author list.
\correspondingauthor{Daniel Carrera}
\email{dcarrera@gmail.com}

\begin{abstract}
% 1) The punchline
We present evidence that it is unlikely that the streaming instability (SI) can form planetesimals from mm grains inside axisymmetric pressure bumps.
% 2) Hi-Res Experiment
We conducted the largest simulation of the SI so far (7 million CPU hours), consisting of a large slice of the disk with mm grains, a solar-like dust-to-gas ratio ($Z = 0.01$), and the largest pressure bump that does not cause gravitational instability (GI) in the particle layer. We used a high resolution of $1000/H$ to resolve as many SI unstable modes as possible. The simulation produced a long-lived particle over-density far exceeding the SI criteria (i.e., a critical solid abundance to headwind parameter ratio $Z/\Pi$) where strong clumping would occur if these conditions were present over an extended region of the disk; yet we observed none. The likely reason is that the time it takes particles to cross the high-$Z/\Pi$ region ($t_{\rm cross}$) is shorter than the growth timescale of the SI ($t_{\rm grow}$). We propose an added criterion for planetesimal formation by the SI --- that $t_{\rm cross} > t_{\rm grow}$.
% 3) Lo-Res Experiment
We show that any bump larger than the one in this run would form planetesimals by the GI instead of the SI.
% 4) Implications
Our results significantly restrict the pathways to planet formation: Either protoplanetary disks regularly form grains larger than 1~mm, or planetesimals do not form by the SI in axisymmetric pressure bumps. Since bumps large enough to induce the GI are likely Rossby-wave unstable, we propose that mm grains may only form planetesimals in vortices.
\end{abstract}

\keywords{accretion disks -- protoplanetary disks -- planets and satellites: formation}

%%%%%%%%%%%%%%%%%%%%%%%%%%%%%%%%%%%%%%%%%%%%%%%%%%
%
%   INTRODUCTION
%
%%%%%%%%%%%%%%%%%%%%%%%%%%%%%%%%%%%%%%%%%%%%%%%%%%
\section{Introduction}
\label{sec:intro}

Ever since its first iconic images of the circumstellar dust rings around HL Tau \citep{Alma_2015}, ALMA has revolutionized our understanding of planet formation. Before ALMA there had been plenty of speculation that gas pressure bumps might play a role in planet formation by concentrating dust grains \citep[e.g.,][]{Johansen_2009,Sandor_2011,Pinilla_2012,Simon_2014,Bai_2014}. However, ALMA has shown that axisymmetric dust rings, and the pressure bumps that likely form them, not only exist, but are ubiquitous in young protoplanetary disks \citep{Huang_2018}. 

These dust rings are a natural site for planetesimal formation. In \citet{Carrera_2021,Carrera_2022} we used 3D simulations (particles + gas) to show that even a fairly small pressure bump can accumulate cm-sized dust grains sufficiently to trigger planetesimal formation. However, it is notable that the particle sizes observed by ALMA are closer to the mm-size range \citep{Huang_2018}. This observation might merely indicate that larger grains have been effectively converted into planetesimals, and there is indeed some tentative evidence that that might be the case, since planetesimal formation seems to explain the relatively uniform optical depths in dust rings \citep{Stammler_2019}. Or the maximum grains size may truly be $<$~mm --- observations of polarized and continuum dust emission from Ophiuchus IRS 48 are consistent with either an optically thin disk with both sub-mm and 5cm grains, or an optically thick disk with sub-mm grains only \citep{Ohashi_2020}. Thus, it is critically important that we understand whether or how planetesimal formation can occur from dust grains no larger than $\sim 1$~mm.  This paper serves that need.

% ---------------------------------------- %
% INTRODUCTION > BARRIERS
% ---------------------------------------- %
\subsection{Barriers to Planet Formation}
\label{sec:intro:barriers}

Planet formation occurs in the circumstellar disks that surround young stars. It begins with the coagulation of micrometer-sized dust particles into larger grains. However, this growth is quickly impeded by numerous growth barriers. As particles grow larger, their collision speeds increase \citep{Weidenschilling_1984,Ormel_2007}. At some point, at around $\sim$mm-cm size, collisions between particles are more likely to lead to bouncing or fragmentation rather than growth \citep{Blum_2008,Guttler_2010,Zsom_2010}. In addition, particles also experience a headwind because the gas component of the disk orbits at a slightly sub-Keplerian speed thanks to its radial pressure support. The difference $\Delta v$ between the Keplerian speed $\vk$ of the particles and the azimuthal gas velocity $u_\phi$ is

\begin{eqnarray}
\Delta v &\equiv& \vk - u_\phi = \eta\vk\\
\eta &\equiv& -\frac{1}{2}\left(\frac{\cs}{\vk}\right)^2
        \frac{\partial \ln P}{\partial \ln r}
\end{eqnarray}

\noindent
where $\cs$ is the sound speed and $P$ is the gas pressure \citep{Nakagawa_1986}. This headwind induces radial drift on the particles due to aerodynamic drag. The rate of radial drift is

\begin{equation}
    u_{\rm drift} = \frac{-2\Delta v}{\tau + \tau^{-1}}
\end{equation}
 \citep{Weidenschilling_1977} where the $\tau$ is the \textit{Stokes number}, or the particle stopping time in units of the Keplerian frequency $\Omega$

\begin{equation}\label{eqn:stokes}
\tau = t_{\rm stop}\Omega
    = \frac{\rhos a}{\rho \cs}\sqrt{\frac{\pi}{8}}\Omega
\end{equation}
where $a$ is the particle size, $\rho$ is the gas density, and $\rhos$ is the density of the solid material (e.g., $\rho_s = 1 {\rm g}\,{\rm cm}^{-3}$ for ice). For $\tau \sim 1$, radial drift itself limits the particle size \citep{Weidenschilling_1977}. This radial drift barrier is generally dominant in the outer disk \citep{Birnstiel_2012}.

% ---------------------------------------- %
% INTRODUCTION > OVERCOMING BARRIERS
% ---------------------------------------- %
\subsection{Overcoming Barriers: Streaming Instability}
\label{sec:intro:overcoming}

Perhaps the biggest open question in planet formation is how dust particles overcome the fragmentation, bouncing, and radial drift barriers to form planetesimals. Planetesimals are the $1-100$ km bodies that are thought to be the building blocks of planets. The most promising formation scenario is that planetesimals form through some form of gravitational instability (GI) of small mm-cm grains. For example, \citet{Goldreich_1973} suggested that solids might sediment into a very thin dust layer at the midplane that would then become gravitationally unstable. However, it was soon realized that the dust layer would be Kelvin-Helmholtz unstable and the turbulence induced by the Kelvin-Helmholtz instability would impede further sedimentation \citep{Weidenschilling_1980}.

Today, most models of planetesimal formation rely on some aerodynamic process to concentrate particles in a small region until the local particle density $\rhop$ exceeds the Roche density

\begin{equation}
\rho_{\rm roche} = \frac{9\Omega^2}{4\pi G},
\end{equation}
so that the dust becomes gravitationally unstable. Some prominent examples include the accumulation of dust in axisymmetric pressure bumps in the disk \citep[e.g.,][]{Taki_2016,Dullemond_2018,Stammler_2019,Carrera_2021}, MHD zonal flows \citep[e.g.,][]{Dittrich_2013,Xu_2021}, disk photo-evaporation \citep[e.g.,][]{Throop_2005,Carrera_2017}, and trapping dust inside vortices \citep[e.g.,][]{Tanga_1996,Cuzzi_2001,Heng_2010,Raettig_2021}. However, the most promising mechanism seems to be the runaway convergence of radial drift known as the \textit{streaming instability} \citep[SI;][]{Youdin_2005,Johansen_2007b}. One important piece of evidence in favor of the SI is that the SI seems to reproduce the inclination distribution of binaries in the Kuiper Belt remarkably well \citep{Nesvorny_2019}.

One of the most important things to know about the SI is that it  only seems to be an efficient process when the dust-to-gas ratio $Z = \Sigmap/\Sigmag$ has already reached some minimum threshold. The SI is most efficient for $\tau \approx 0.1 - 0.5$. For smaller $\tau$, the critical $Z$ needed to produce particle densities in excess of $\rho_{\rm Roche}$ rapidly increases to $Z \gg 0.01$ \citep{Carrera_2015,Yang_2017,Li_2021}. The next thing to know about the SI is that it is more efficient when the headwind $\Pi = \Delta v/\cs$ induced by the pressure gradient is lower \citep{Bai_2010b,Carrera_2015}. In fact, \citet{Sekiya_2018} showed that the overall structure of the particle filaments produced by the SI seems to scale with $Z/\Pi$. This result allows us to re-scale the strong clumping criteria of \citet{Carrera_2015,Yang_2017,Li_2021} for any pressure profile (see Section~\ref{sec:results}). Finally, there are theoretical predictions \citep{Chen_2020,Umurhan_2020} and numerical models \citep{Gole_2020} that indicate that the SI may have trouble forming planetesimals in turbulent disks, though that seems to depend on the type of turbulence \citep{Yang_2018}. A proper study of turbulence is beyond the scope of this work; our runs only have a small amount of turbulence generated by particle streaming at the midplane.

It is quite possible that two or more mechanisms (the SI, pressure bumps, zonal flows, etc.) work together in tandem. After \citet{Carrera_2015} identified the $Z-\tau$ dependence of the SI, several authors have sought to combine the SI with some other process that also concentrates particles, such as pressure bumps \citep{Auffinger_2018,Stammler_2019,Carrera_2021,Carrera_2022},  snow lines \citep{Drazkowska_2017}, disk photo-evaporation \citep{Carrera_2017}, vortices \citep{Regaly_2021}, and MHD zonal flows \citep{Xu_2021}.

In this work we follow up on the results of \citet{Carrera_2021,Carrera_2022} who found that, for cm-size particles, pressure bumps reliably create the conditions necessary to trigger planetesimal formation by the SI. Importantly, the pressure bump does not need to be very large, and a particle trap (meaning that particle drift is halted) is not needed to form planetesimals. However, the results for smaller mm-size particles were inconclusive. This is an important limitation because the maximum particle sizes that we observe in ALMA rings may be $\sim$~mm (see discussion above). Thus, it is important to ascertain whether or not planetesimals can form with mm grains only. In this paper we extend the work of \citet{Carrera_2021,Carrera_2022} with a new high-resolution simulation with mm-size grains and a large pressure bump. Our single objective is to determine whether the SI can form planetesimals in a pressure bump out of mm-size grains.

This paper is organized as follows. In section \S\ref{sec:methods} we describe our numerical methods, including our disk model (\S\ref{sec:methods:disk}), pressure bump model (\S\ref{sec:methods:bump}), and our simulation setup (\S\ref{sec:methods:setup}). Our results are presented in section \S\ref{sec:results}. In section \S\ref{sec:discussion} we propose a way to interpret our results and suggest a new planetesimal formation criterion. Finally, our summary and conclusions are found in section \S\ref{sec:conclusions}.

%%%%%%%%%%%%%%%%%%%%%%%%%%%%%%%%%%%%%%%%%%%%%%%%%%
%
%   METHODS
%
%%%%%%%%%%%%%%%%%%%%%%%%%%%%%%%%%%%%%%%%%%%%%%%%%%
\section{Numerical methods}
\label{sec:methods}

We employ identical numerical methods as \cite{Carrera_2021}, with different parameters. Readers who are familiar with our previous works may prefer to read the following short text, skip the rest of this section, and continue on to Section~\ref{sec:methods:setup}: We conducted local, shearing box simulations with the {\sc Athena} code that include gas and particles but have no magnetic fields or externally driven turbulence. The gas is treated as a compressible, isothermal fluid, and the particles are Lagrangian super-particles. Particle self-gravity is implemented using a particle-mesh approach with shear-periodic radial boundary conditions. In the rest of this section we summarize our methods. We direct the reader to \cite{Carrera_2021} for additional details.

%------------------------------%
% METHODS > HYDRO
%------------------------------%
\subsection{Hydrodynamic Solver}
\label{sec:methods:hydro}

We use the {\sc Athena} code \citep{Stone_2008} in pure hydrodynamic mode with particle feedback and no magnetic fields and an isothermal equation of state $P = \rho\cs^2$. We use the shearing local shearing box approximation, where a patch of the disk is treated as a local Cartesian frame $(x,y,z)$, neglecting disk curvature. The local frame is defined in terms of the disk's cylindrical coordinates $(R,\phi,z^\prime)$ as

\begin{eqnarray*}
  x &=& (R-R_0)\\
  y &=& R_0 \phi\\
  z &=& z^\prime
\end{eqnarray*}
where $R_0$ is the center of the box. {\sc Athena} includes a super-particle approach in which each super-particle is a statistical representation of many smaller particles. All of our runs have the same number of super-particles as there are grid cells. Super-particle $i$ is governed by an equation of motion:

\begin{eqnarray}
  \label{eqn:particle_motion}
  \frac{d {\bmath v^\prime_i}}{dt} = 2\left( v^\prime_{iy} - \eta \vk \right)& & \Omega \hat{\bmath x} - \left(2 - q\right) v^\prime_{ix} \Omega \hat{\bmath y} \nonumber \\ 
& & - \Omega^2 z \hat{\bmath z} - \frac{{\bmath v^\prime_i} - {\bmath u^\prime}}{\tstop} + {\bmath F_{\rm g}}.
\end{eqnarray}

\noindent
The prime denotes a frame in which the background shear velocity has been subtracted. The $-2 \eta\vk\Omega \hat{\bmath x}$ term is responsible for inward radial drift due to aerodynamic drag. The ${\bmath F_{\rm g}}$ term represents particle self-gravity. The size of the timestep is set by the Courant condition, and is typically $\sim 3\times 10^{-4}\Omegainv$ for a $1000/H$ resolution. We use the same methods as \cite{Simon_2016} to solve the Poisson equation --- we use a triangular shaped cloud (TSC) scheme to map the mass density of particles to grid cells and use a Fast Fourier Transform to solve the Poisson equation,

\begin{eqnarray}
    \label{eqn:poisson}
    \nabla^2 \Phi &=& 4\,\pi\,G\,\rhop. \\
    {\bmath F_{\rm g}} &=& - \nabla \Phi
\end{eqnarray}
We calculate ${\bmath F_{\rm g}}$ by a central finite difference and then interpolate to the locations of the particles via TSC. For more details we refer the reader to \cite{Simon_2016}.

%------------------------------%
% METHODS > BOUNDARY CONDITIONS
%------------------------------%
\subsection{Boundary Conditions}
\label{sec:methods:bc}

The boundary conditions are the same for the gas and particle components: shearing-periodic in the radial dimension \citep{Hawley_1995a}, periodic in azimuth, and a modified outflow boundary in the vertical dimension in which gas density is extrapolated into the ghost zones using an exponential function \citep{Simon_2011,Li_2018}. The modified outflow boundary condition will not entirely prevent gas mass loss along the vertical boundary. To ensure that mass is globally conserved, we renormalize the gas density in every cell at every time step to keep the total gas mass constant. As for the particles, we verify that no particles escape the simulation box through the vertical boundaries. \citet{Li_2018} tested the effect of different vertical boundary conditions (vBC), and found that the outflow vBC produced less stirring of the particle layer for small boxes ($L_z = 0.2H$), but for a large box such as ours ($L_z = 0.8H$) all their choices of vBC produced essentially the same results.

The gravitational potential has the same boundary conditions in the radial and azimuthal directions as the gas and particles. However, the vertical boundary conditions are open with the potential in the ghost zones calculated via a third order extrapolation.

%------------------------------%
% METHODS > DISK MODEL
%------------------------------%
\subsection{Disk Structure}
\label{sec:methods:disk}

The simulation box is centred at $r = 50$AU for a disk with surface density $\Sigma$ and temperature $T$ given by

\begin{eqnarray}
    \Sigma(r) &=& \frac{M_{\rm disk}}{2\pi r_c} r^{-1}\\
    T(r) &=& 280 \left( \frac{r}{{\rm AU}} \right)^{-1/2} {\rm K}
\end{eqnarray}
where $M_{\rm disk} = 0.09 M_\odot$ and $r_c = 200$AU. The temperature corresponds to an optically thin disk. The stellar mass is set to $M_\star = 1 M_\odot$. The disk is slightly flared, with aspect ratio

\begin{equation}
    \frac{H}{r} = \frac{\cs}{\vk} = 0.033 \left( \frac{r}{{\rm AU}} \right)^{1/4}
\end{equation}

\noindent
where $\cs$ is the isothermal sound speed and $\vk$ is the Keplerian orbital speed. We begin with a vertically stratified Gaussian gas density profile. The gas density is uniform along $(x,y)$ apart from small random initial perturbations. The initial concentration of solids is

\begin{equation}
 Z \equiv \Sigma_p/\Sigmag = 0.01   
\end{equation}
with an initial particle scale height of $H_p = 0.025 H$. We fix the particle size to $a \equiv 1$~mm and compute the particle Stokes number $\tau$ dynamically (Equation \ref{eqn:stokes}). The background pressure gradient (i.e., not accounting for the pressure bump) is

\begin{equation}
    \label{eqn:PI}
    \Pi \equiv \frac{\Delta v}{\cs}
        = - \frac{1}{2} \left( \frac{\cs}{\vk} \right) \frac{d \ln P}{d \ln r} = 0.12 
\end{equation}
where $\Delta v$ is the headwind experienced by solid particles. The strength of self-gravity is set by $4\pi G \rho_0 \approx 0.2\Omega^2$ which corresponds to a Toomre \citep{Toomre_1964} $Q$ value of $Q \approx 8$.

%------------------------------%
% METHODS > PRESSURE BUMP
%------------------------------%
\subsection{Pressure Bump}
\label{sec:methods:bump}

Starting with the background gas profile described in the previous section, we gradually form a pressure bump with a Gaussian density profile centered on $x = 0$ with width $w = 1.14H$

\begin{equation}
  \label{eqn:rho_bump}
  \hat{\rho}(x,y,z) = \rho_0 \left[ 1 + A e^{\left(-x^2/2w^2\right)}\right] e^{\left(-z^2/2H^2\right)}.
\end{equation}

Given $A$, the azimuthal speed that maintains geostrophic balance is

\begin{equation}
  \label{eqn:uy_bump}
  \hat{u}_y(x,y,z) = \frac{-A x\cs^2e^{\left(-x^2/2w^2\right)}}{ 2w^2\Omega\left[ 1 + A e^{\left(-x^2/2w^2\right)}\right]}
\end{equation}

In all our simulations the bump amplitude starts at zero and grows gradually. To develop and sustain the pressure bump we use the Newtonian relaxation scheme of \citet{Carrera_2021}. At each timestep $\Delta t$ we adjust $\rho$ and $u_y$ by

\begin{eqnarray}
  \label{eqn:T_reinf}
  \Delta \rho &=& (\hat{\rho} - \rho)  \frac{\Delta t}{\treinf} \\
  \Delta u_y  &=& (\hat{u}_y  - u_y )  \frac{\Delta t}{\treinf}
\end{eqnarray}
where $\treinf = 1\Omegainv$ is the reinforcement timescale. The Newtonian relaxation scheme is maintained for the duration of the simulation

%%%%%%%%%%%%%%%%%%%%%%%%%%%%%%%%%%%%%%%%%%%%%%%%%%
%
%   SIMULATION SETUP
%
%%%%%%%%%%%%%%%%%%%%%%%%%%%%%%%%%%%%%%%%%%%%%%%%%%
\subsection{Simulation Setup}
\label{sec:methods:setup}

Our experiment consists of only three simulations:

\begin{description}
  \item[\texttt{high-res,A=0.5}] \hfill \\
    Our main experiment is a high-resolution (1000 zones/H) simulation with mm-size particles and a Gaussian pressure bump with amplitude $A = 0.5$. This is just below the amplitude that would lead to a particle trap ($dP/dr = 0$).
  \item[\texttt{low-res,A=0.5} and \texttt{low-res,A=0.6}] \hfill \\
    Two much lower resolution runs (160 zones/H) with a pressure bump amplitudes of $A = 0.5,0.6$. The low resolution is intended to \textit{not} resolve the fastest growing modes of the SI, and demonstrate that $A = 0.6$ is just enough to form planetesimals by GI.
\end{description}

\noindent
In practice, we conducted several \texttt{low-res} runs with different pressure bump amplitudes to find the critical value where \texttt{low-res} started to form planetesimals by the GI. That turned out to be $A_{\rm GI} = 0.6$. That became \texttt{low-res,A=0.6}. We then conducted a \texttt{high-res} for a pressure bump amplitude just a step below $A_{\rm GI}$, and that became \texttt{high-res,A=0.5}. Note that a \texttt{high-res} run with $A = A_{GI}$ would almost certainly show the GI as well because higher resolution allows for higher particle densities and better resolves the FFT-based gravitational potential.

\begin{figure}[ht]
    \centering
    \includegraphics[width=1\linewidth]{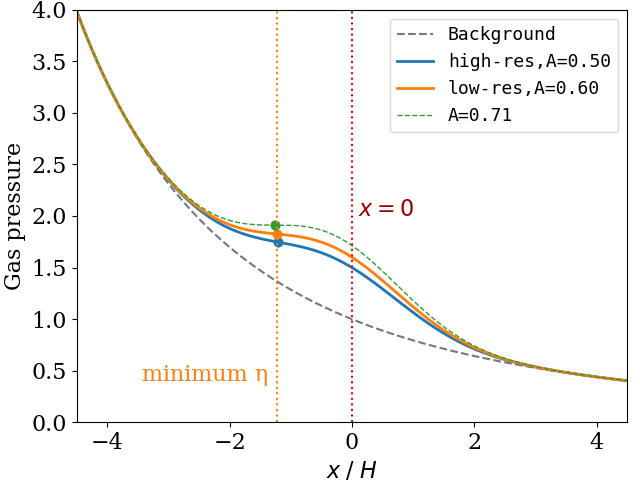}
    \caption{Gas pressure profile with a pressure bump and without. Our simulations have bump amplitudes of $A = 0.5$ (\texttt{high-res}, blue) and $A = 0.6$ (\texttt{low-res}, orange). For reference, we also show the pressure profile without a bump (gray) and with the smallest bump with a particle trap ($\max[dP/dr] \ge 0$; green). Because the pressure bump is sitting on top of a background pressure gradient, the point of minimum headwind is offset from the center of the simulation box.}
    \label{fig:density_profile}
\end{figure}

In all cases, the simulation domain is a shearing box with dimensions $L_x\times L_y\times L_z = 9H\times0.2H\times0.8H$, where $H$ is the gas scale height. Figure \ref{fig:density_profile} shows the pressure density profile for our three simulations ($A = 0.5$ and $0.6$), as well as the background pressure profile. For reference, we also show the profile for the smallest pressure bump that has a true particle trap. Notice also that neither of our three runs have a local pressure maximum, and thus neither have a particle trap. In addition, the point of minimum headwind, where the SI should be most effective, depends weakly on $A$ and is offset from the center of the box (at around $x = -1.2H$).

Lastly, the initial particle distribution is horizontally uniform but normally distributed along the vertical axis, with a particle scale height of $H_p = 0.025H$.  To seed the SI, the positions of the particles are given small random perturbations away from this distribution.

%%%%%%%%%%%%%%%%%%%%%%%%%%%%%%%%%%%%%%%%%%%%%%%%%%
%
%   RESOLUTION
%
%%%%%%%%%%%%%%%%%%%%%%%%%%%%%%%%%%%%%%%%%%%%%%%%%%
\subsection{Resolution and the SI}
\label{sec:methods:resolution}

Our key goal is to determine whether the SI can form planetesimals for mm-size particles. Proving that the SI \textit{is not} active is challenging because the fastest growing modes of the SI are difficult to resolve. Our high-res run is intended to do just this to the greatest extent possible. The resolution needed to resolve the SI is notoriously difficult to estimate, especially in the non-linear regime. Our best guide is the numerical work of \cite{Yang_2017}, who saw SI filaments for $\tau = 0.01$ and a resolution of 640/H. Compared to that work, our runs have higher resolution (1000/H) and larger particles ($\tau \approx 0.0123$). \citet{Yang_2017} found that higher resolution and larger particle sizes both cause filaments to form earlier. They also found that the largest voids in streaming turbulence scale with $\tau$. This suggests that the modes of the non-linear regime (and hence, the grid size required) scale with $\tau$, consistent with the linear regime. For these reasons, we feel confident that \texttt{high-res,A=0.50} is sufficiently well resolved to capture the SI.

We adopt a higher resolution than \cite{Yang_2017} in order to give our \texttt{high-res} run every opportunity to grow the SI as quickly as possible. Because there is no theoretical estimate of the fastest growing wavelength in the non-linear regime and a turbulent medium, we are limited to a back-of-the-envelop estimate based on linear theory in a laminar environment: Within the ``resonant-drag instability'' regime of the SI, the fastest growing modes are those that satisfy the epicyclic resonant condition \citep{Squire_2020}

\begin{equation}
    \label{eqn:si_fast}
    \mathbf{k} \cdot \mathbf{w}_s = \hat{k}_z \Omega,
\end{equation}
where $\mathbf{k} = (k_x, 0, k_z)$ is the wavenumber, $\mathbf{w}_s$ is the dust drift velocity with respect to the gas, and $\hat{k}_z = k_z/k$. \citet{Nakagawa_1986} showed that $\mathbf{w_s} \approx -2 \tau \eta \; \vk \; \mathbf{\hat{x}}$. Therefore, the fastest growing mode has wavelength

\begin{equation}
    \label{eqn:lambda}
    \lambda \approx -4\pi\tau \eta\;r\;\frac{k_x}{k_z}.
\end{equation}

\noindent
For the sake of simplicity let us assume that $-k_x/k_z \approx 1$. Given our typical particle Stokes number of $\tau = 0.0123$ for mm-size particles and bump amplitude, we compute $\lambda$ numerically across the simulation domain.

\begin{figure}
    \centering
    \includegraphics[width=0.47\textwidth]{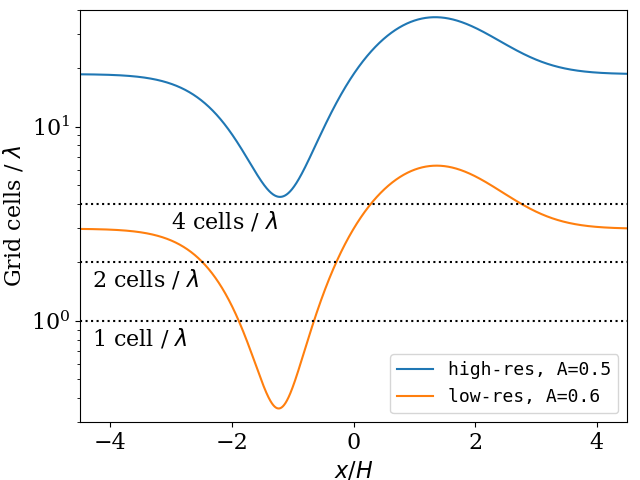}
    \caption{Number of grid cells per $\lambda$ for our two resolutions, where $\lambda$ is the wavelength of the fastest growing mode of the SI (equation~(\ref{eqn:lambda})). The high resolution run has just enough resolution to resolve the fastest growing modes everywhere, despite the small particle size and low $\eta$ inside the pressure bump. The low resolution run does not adequately resolve the SI in the planetesimal formation region, as the number of cells per $\lambda$ is $< 1$.}
    \label{fig:resolution}
\end{figure}

Figure \ref{fig:resolution} shows the number of grid cells per $\lambda$ across the simulation for our \texttt{high-res} and \texttt{low-res} runs. The figure shows that our \texttt{high-res} run can resolve the fastest growing mode of the SI everywhere with at least four grid cells per $\lambda$. Conversely, the \texttt{low-res} run quite intentionally \textit{cannot} resolve the fastest growing modes of the SI in the planetesimal formation region. Therefore, if this run forms planetesimals it would potentially be through a non-SI process (see Section \ref{sec:results}).

%%%%%%%%%%%%%%%%%%%%%%%%%%%%%%%%%%%%%%%%%%%%%%%%%%
%
%   RESULTS
%
%%%%%%%%%%%%%%%%%%%%%%%%%%%%%%%%%%%%%%%%%%%%%%%%%%
\section{Results}
\label{sec:results}

Our key results are quite simple:

\begin{itemize}
\item The \texttt{high-res,A=0.5} run never forms planetesimals, or even gets close to the Roche density, despite its high resolution and long runtime.

\item The \texttt{low-res,A=0.6} run forms planetesimals and \texttt{low-res,A=0.5} does not. Furthermore, we can show that the planetesimals in \texttt{low-res,A=0.6} form exclusively by the GI.
\end{itemize}

\begin{figure}
    \centering
    \includegraphics[width=0.47\textwidth]{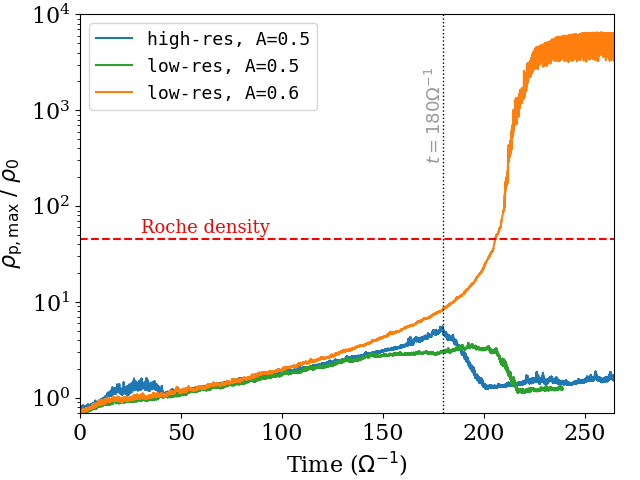}
    \caption{Maximum particle density, normalized to the initial mid-plane gas density, for all runs. \textit{Blue:} $A = 0.5$ cannot form planetesimals despite having enough resolution to resolve the fastest growing modes of the SI. \textit{Green and Orange:} $A = 0.6$ forms planetesimals, even at low resolution, but $A = 0.5$ does not. The vertical line marks the time of the snapshot shown in Figure \ref{fig:hi-res-1800}.}
    \label{fig:dmax}
\end{figure}

\begin{figure}
    \centering
    \includegraphics[width=0.47\textwidth]{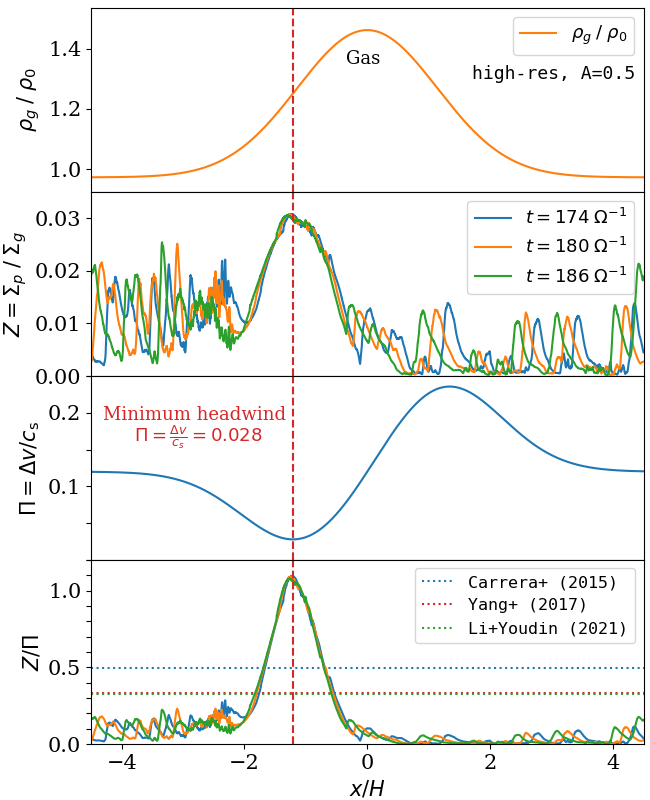}
    \caption{\textit{1st Plot:} Gas density profile at the midplane, normalized to the initial mid-plane gas pressure, across the simulation domain for \texttt{high-res,A=0.5}. \textit{2nd Plot:} Dust-to-gas ratio for three snapshots in the vicinity of $t = 180\Omegainv$. Wide particle filaments form everywhere, and a steady-state particle pileup occurs in the region of minimum headwind $\eta$. The dust-to-gas ratio peaks at $Z \gtrsim 0.03$ with a local headwind of $\Pi \approx 0.0283$. \textit{3rd Plot:} Headwind parameter $\Pi$. \textit{4th Plot:} Normalized dust-to-gas ratio $Z/\Pi$, along with the SI criteria (dotted lines). These criteria are also shown in Figure \ref{fig:SI_criterion}.}
    \label{fig:steps}
\end{figure}

Figure \ref{fig:dmax} shows the maximum particle density for all of our runs as a function of time. For \texttt{high-res, A=0.5} the density peaks at $t = 180\Omegainv$ (vertical dotted line) at only 12\% of the Roche density. Figures \ref{fig:steps} and \ref{fig:hi-res-1800} show what is happening in the simulation at around this time. Figure \ref{fig:steps} shows the dust-to-gas ratio $Z = \Sigma_p / \Sigma_g$ for three snapshots centered around $t = 180\Omegainv$. Particle feedback forms large scale particle filaments across the entire domain, but we note that these are much wider and less dense structures than the ``strong clumping'' associated with the SI \citep[$\sim 0.3H$ vs $\sim 0.01H$][]{Yang_2017}.

\begin{figure}
    \centering
    \includegraphics[width=0.47\textwidth]{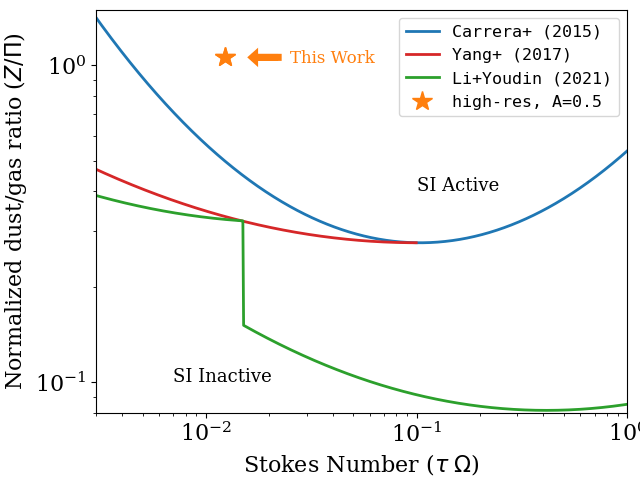}
    \caption{Solid abundance criterion for strong clumping in terms of $Z/\Pi$ vs $\tau$ as reported in \citet[blue]{Carrera_2015}, \citet[red]{Yang_2017}, and \citet[green]{Li_2021}, but for the case without a pressure bump. The orange star marks the $(\tau, Z/\Pi)$ for the \texttt{high-res,A=0.5} snapshots in Figure \ref{fig:steps} ($\tau \approx 0.0123$, $Z/\Pi \approx 1.06$). If the conditions in the density peak in our \texttt{high-res} run were applied over an extended radial region of the disk, previous works show that the SI would produce strong clumping.}
    \label{fig:SI_criterion}
\end{figure}

\begin{figure*}
    \centering
    \includegraphics[width=0.98\textwidth]{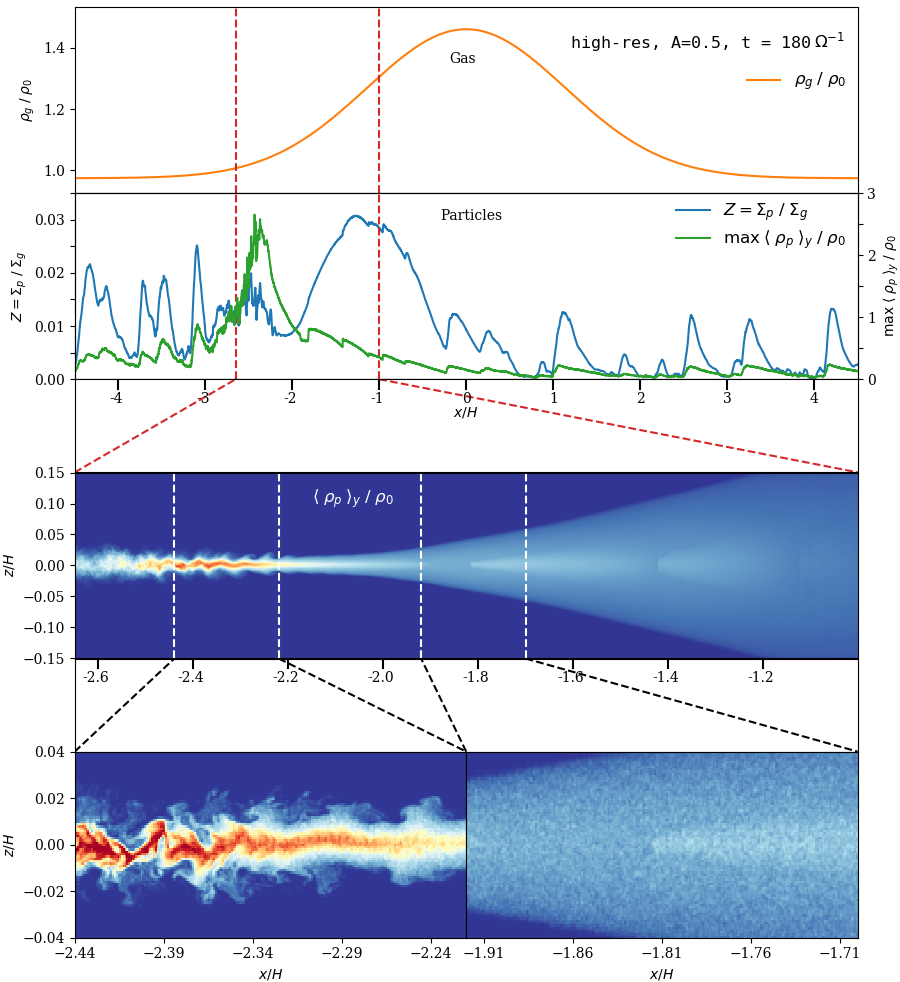}\\
    \includegraphics[width=0.45\textwidth]{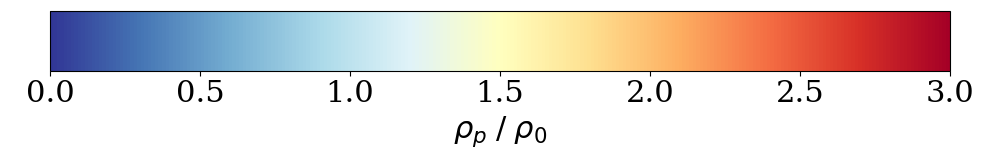}
    \caption{Snapshot of \texttt{high-res, A=0.5} at $t = 180\Omegainv$, when the maximum particle density is reached (Figure \ref{fig:dmax}). The third plot shows the mean particle density; $\rho_0$ is the initial mid-plane gas density at $x = \pm\infty$. The bottom plots zoom in further to two regions that show interesting structure. They show a \textit{slice} of the simulation at $y = 0$. The left region shows features of Kelvin-Helmholtz turbulence. The peak density at the midplane is a temporary aberration; as turbulence develops the density drops (Figure \ref{fig:hi-res-2647}). The right region shows a small density peak, but little structure. The local pressure gradient is $\Pi \approx 0.028$.}
    \label{fig:hi-res-1800}
\end{figure*}

While particle filaments drift across the simulation, Figure \ref{fig:steps} also shows a large steady-state particle pileup around the point with minimum headwind $\eta$. There, the dust-to-gas ratio reaches $Z \gtrsim 0.03$ at its peak. \citet{Sekiya_2018} showed that the clumping structure of the SI scales with $Z/\Pi$. In the region within $\pm 0.1H$ from the peak in $Z$ has $0.028 < \Pi < 0.030$, which gives a normalized dust-to-gas ratio of $Z/\Pi \approx 1.06$. For comparison, we also show the strong clumping criteria obtained by \citet{Carrera_2015,Yang_2017,Li_2021} for this particle size. In addition, Figure \ref{fig:SI_criterion} shows these criteria in terms of $Z/\Pi$ vs $\tau$. The key take-away is that our \texttt{high-res} simulation has a very prominent peak in $Z/\Pi$ that is deep inside the region where the SI would cause strong clumping if the conditions held over an extended radial region of the disk. Yet, our \texttt{high-res} simulation shows no sign of strong clumping. Furthermore, the simulation has run long enough ($t_{\rm max} = 265\Omegainv$) for filaments to develop \citep[Figure 2 of][]{Li_2021}.

The difference in local conditions between our runs and those of \citet{Carrera_2015,Yang_2017,Li_2021} reflects a limitation of all small-box simulations. They all effectively assume that the particular value of $Z$ is maintained over an extended region of the disk, which may not translate well to pressure bumps where high $Z$ is only attained in a small region. Only a high-res large-box simulation like ours can determine whether localized changes to $(Z,\Pi)$ lead to planetesimal formation.

\begin{figure*}
    \centering
    \includegraphics[width=0.98\textwidth]{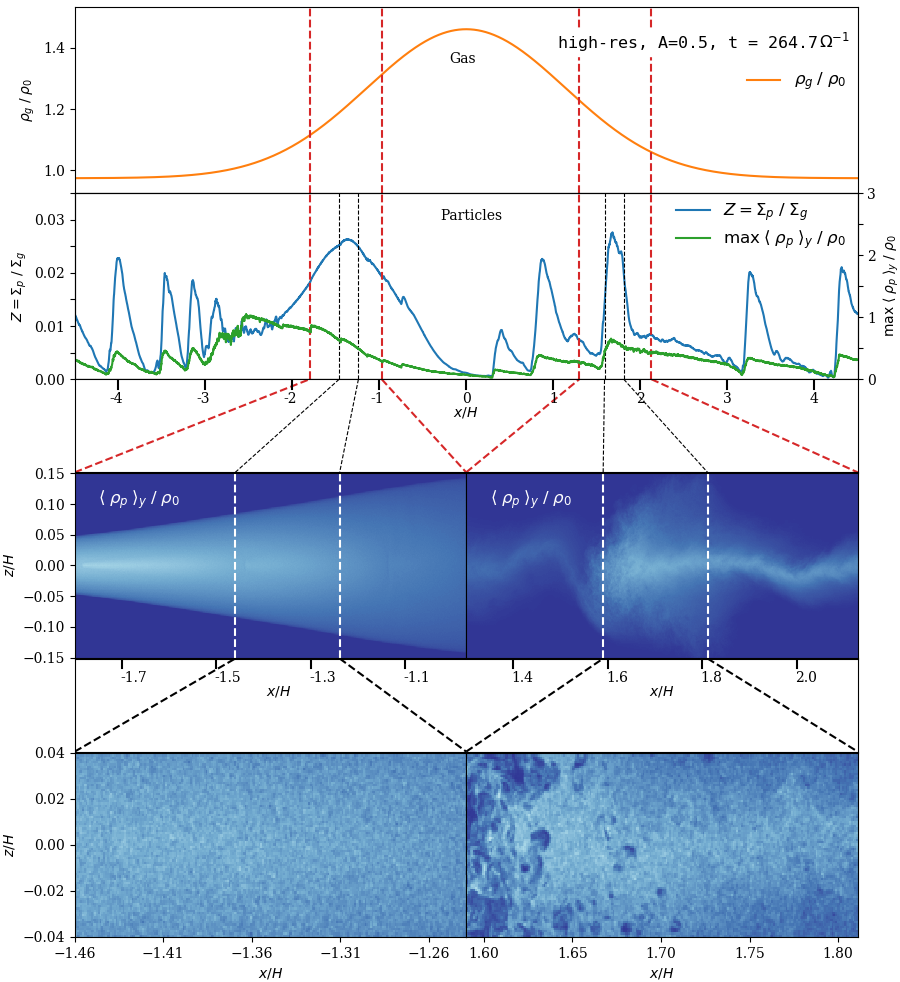}\\
    \includegraphics[width=0.45\textwidth]{colorbar_horizontal.png}
    \caption{Snapshot of \texttt{high-res,A=0.5} at $t = 264.7\Omegainv$, which is our final snapshot and it is later than the time when Run Z2T1 of \citet{Li_2021} formed filaments. The third row shows the mean particle density; $\rho_0$ is the initial mid-plane gas density. The last row zooms in further to the regions with the highest $Z$. They show a \textit{slice} of the simulation at $y = 0$. Neither region shows strong clumping. For the region on the right, $Z \approx 0.027$ and $\Pi \approx 0.226$, so the lack of strong clumping is expected. But on the left, $Z \approx 0.025$ and $\Pi \approx 0.029$ is well inside the region where strong clumping would occur if the conditions present in the pressure bump were met over a wider radial region of the disk.}
    \label{fig:hi-res-2647}
\end{figure*}

Figure \ref{fig:hi-res-1800} shows the snapshot at $t = 180\Omegainv$ in greater detail. The density maximum is not inside the particle pileup; instead, it occurs about $\sim 1.1H$ downstream, at $x \approx -2.4H$, and it merely reflects a brief moment when the particles have sedimented to the midplane but turbulence has not fully developed. At $t = 180\Omegainv$ one can see features of Kelvin-Helmholtz turbulence around $-2.5H < x < -2.2H$. Using the convention employed by \citet{Gole_2020}, we calculate that $\alpha = 6\times 10^{-5}$ in the active region and near the midplane ($|z| < 0.02H$), compared to an average of $\alpha = 4\times 10^{-6}$ across the entire domain. At later times we see similar features but with a higher particle scale height. After $t \approx 200\Omegainv$ we seem to reach a steady state with no sign of strong clumping (see Fig.~\ref{fig:dmax}).

We estimated the total rate of mass loss through the vertical outflow boundaries. This simulation loses around $\Delta m/m < 10^{-6}$ of its total mass per timestep. Most of the outflow occurs in front of the particle filaments. For example, for the snapshot shown in Figure \ref{fig:hi-res-1800}, the rate of outflow ($\Delta m/m = 4\times 10^{-6}$) occurs at $x = 1.04H$. The active region ($-2.5H < x < -2.2H$) actually has one of the lowest rates of outflow ($\Delta m/m = 3\times 10^{-10}$), as the turbulence is heavily concentrated at the midplane.

Figure \ref{fig:hi-res-2647} shows our final snapshot, at $t = 264.7\Omegainv$. This is later than the time when \citet{Li_2021} saw clear evidence of filament formation for $Z = 0.02, \Pi = 0.05, \tau = 0.01$. At this point the simulation has had time to develop turbulence and seems to have reached something close to a steady state. The value of $Z/\Pi$ has decreased slightly to $\approx 0.88$ relative to the snapshots in Figure \ref{fig:steps}, but that value is still $\approx 2.7$ times larger than the value where \citet{Li_2021} found strong clumping under the assumption that these conditions were held over an extended radial region of the disk. In other words, if those results were a predictor of the behavior of the SI in pressure bumps, the SI should be easily visible, yet there is none. This shows that it is difficult to trigger the SI for small particles when high $Z/\Pi$ is only attained in a small region. In section \S\ref{sec:discussion:residency} we discuss the most likely reason why strong clumping failed.

For the sake of completeness, Figure \ref{fig:hi-res-2647} also shows the structure in the densest particle filament (at $x \approx 1.7H$), though in that case we do not expect strong clumping because the $Z/\Pi$ ratio is quite low.

\begin{figure*}
    \centering
    \includegraphics[width=0.98\textwidth]{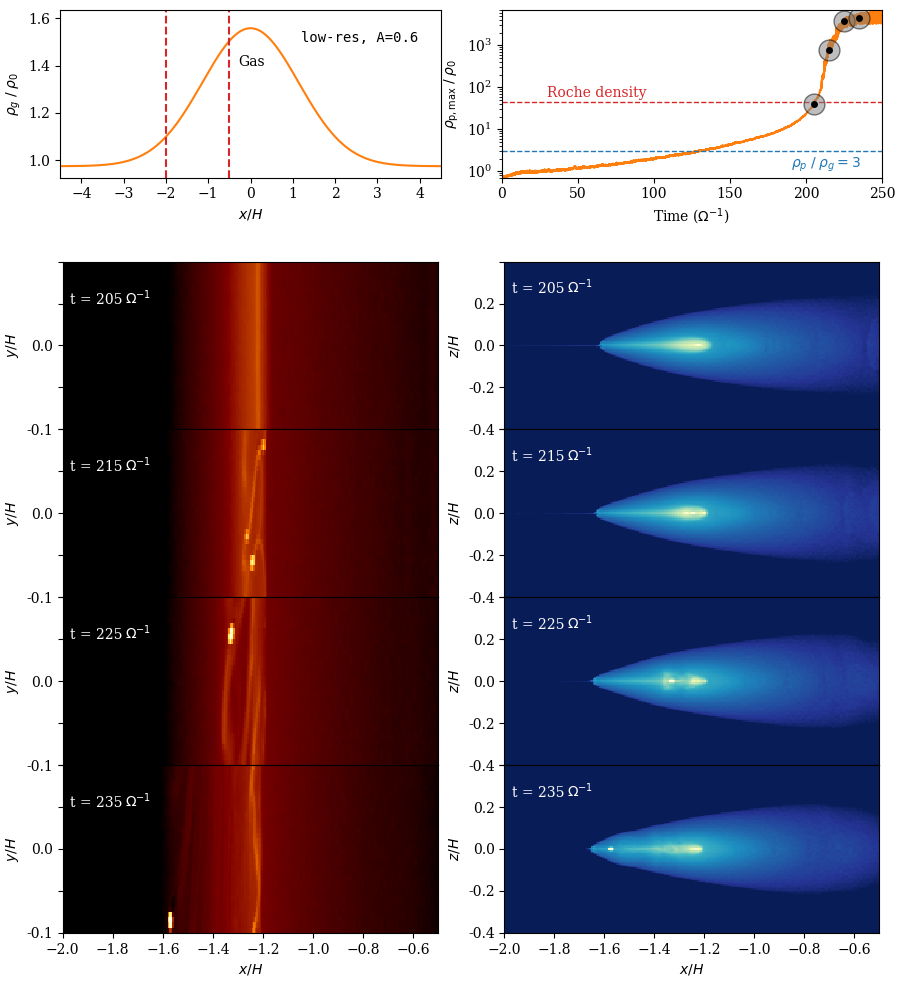}\\
    \hspace{7.7mm}
    \includegraphics[width=0.455\textwidth]{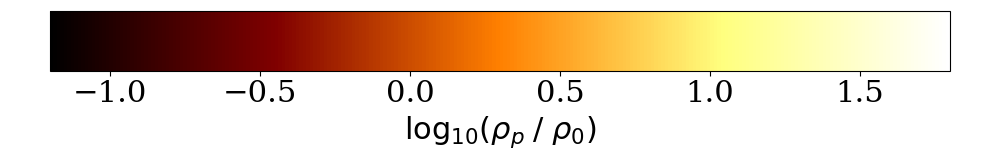}
    \hspace{2.3mm}
    \includegraphics[width=0.455\textwidth]{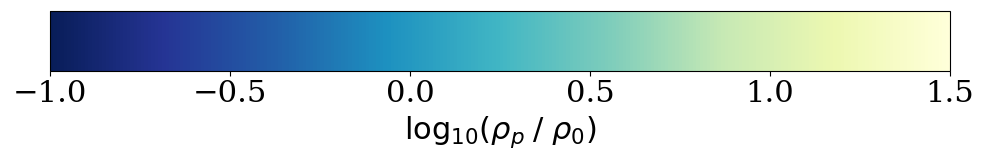}
    \caption{Four snapshots of the \texttt{low-res,A=0.6} simulation starting at the time when the particle density is about to reach the Roche density. The top-left shows the gas density profile, and vertical dashed lines that mark the $-2 \le x/H \le -0.5$ zoom-in region of the snapshots; $\rho_0$ is the initial mid-plane gas density. The top-right shows the maximum particle density over time, with the times of the snapshots marked. The simulation only begins to develop non-axisymmetric structure after it crosses the Roche density --- a tell tale sign of gravitational instability.}
    \label{fig:low-res}
\end{figure*}

Before conducting the \texttt{high-res} run, we conducted several \texttt{low-res} runs at various amplitudes. We found that \texttt{low-res,A=0.5} did not form any structures but \texttt{low-res,A=0.6} did. The purpose of this experiment is to show that any pressure bump much larger than $A = 0.5$ can form GI-unstable particle pileups without the need for the SI. Figure \ref{fig:low-res} shows four snapshots of \texttt{low-res,A=0.6}. The snapshots start at $t = 205\Omegainv$ which is an instant before the maximum particle density $\rho_{p,\rm max}$ reaches the Roche density ($\rho_R/\rho_0 = 45$). The other snapshots are spaced in even intervals of $10\Omegainv$. The snapshots show the azimuthally and vertically averaged particle density in the region $-2H < x < -0.5H$. The top-left plot shows the gas density profile with dashed lines at $x \in \{-2H, -0.5H\}$ for reference. The top-right plot shows the maximum particle density versus time, with black circles marking the times when the snapshots are taken.

The snapshots in Figure \ref{fig:low-res} show that the formation of non-axisymmetric structure exactly coincides with crossing the Roche density. This is a tell tale sign of gravitational instability. This run reaches $\rhop/\rhog = 3$ at $t = 128.5\Omegainv$ (blue dashed line of top-right plot). This is a useful reference point. \citet{Yang_2017} conducted a simulation with the same particle size and resolution as \texttt{low-res} with $\rhop/\rhog \gtrsim 3$ that did not show any clumping after $1,000$ orbits. In contrast, \texttt{low-res,A=0.6} reaches the Roche density after twelve more orbits ($t = 206\Omegainv$). All this is to show that the planetesimal formation in Figure \ref{fig:low-res} occurs purely through the GI: A particle trap concentrates solids until the particle density crosses the Roche density.

Is it possible that a higher resolution run with $A = 0.6$ might have shown signs of the SI? Yes, of course. But our primary point is that the SI is not needed. Taken together, these runs prove one thing: \textit{for mm-size particles, any pressure bump large enough to form planetesimals by the SI is also large enough to form planetesimals without it}.

%%%%%%%%%%%%%%%%%%%%%%%%%%%%%%%%%%%%%%%%%%%%%%%%%%
%
%   DISCUSSION
%
%%%%%%%%%%%%%%%%%%%%%%%%%%%%%%%%%%%%%%%%%%%%%%%%%%
\section{Discussion}
\label{sec:discussion}

\subsection{New SI criterion: Residence time}
\label{sec:discussion:residency}

Considering that the \texttt{high-res} simulation reached high $Z/\Pi$ values, why did it fail to form planetesimals? The most plausible explanation is that each particle's \textit{residence time} inside the high $Z/\Pi$ region is shorter than the SI growth timescale. Once particles finish crossing the density peak, any progress toward SI filaments is essentially forgotten.

This suggests a new criterion for the SI: $t_{\rm cross} > t_{\rm grow}$. Let $\ell$ be the width of a region in a protoplanetary disk with $Z/\Pi$ suitable for planetesimal formation. The particle crossing time is $t_{\rm cross} = \ell / |v_r|$, where $v_r$,

\begin{equation}
v_r = \frac{-2\Pi \cs}{\tau + 1/\tau},
\end{equation}
is the particle drift rate \citep{Weidenschilling_1977}. In the case of our \texttt{high-res} simulation we have $\tau = 0.0123$, and the density peak is roughly $\Delta x \approx 0.1H$ wide with $\Pi \approx 0.0283$. That works out to a nominal crossing time of $t_{\rm cross} \approx 144\Omegainv$. This value is comparable to the time needed for SI filaments to become visible for this particle size \citep{Yang_2017,Li_2021}. One additional complication is that in our simulations, and in any scenario where an external force applies a torque on the gas, the particle drift rate may be affected by that torque as well \citep{Carrera_2022}. In our \texttt{high-res} run the particle drift rate across the $Z/\Pi$ region is $\approx 12$ times higher than in the above estimate. Therefore, the particle residency time in the high $Z/\Pi$ region in \texttt{high-res} is significantly smaller than the growth time of the SI for this particle size.

\subsection{Can the SI win against GI for $A = 0.6$?}
\label{sec:discussion:A06}

In other words, if \texttt{low-res,A=0.6} had instead had a much higher resolution, would we have seen the SI form dense particle filaments before the GI took over? We can use our residence time criterion to make an educated guess. At its minimum, $\Pi = 0.0143$, and within $\pm 0.1H$ we have a mean value of $\Pi \approx 0.0145$. That works out to about half of the drift rate as for $A = 0.5$ and a crossing time of $t_{\rm cross} \approx 280\Omegainv$, before taking into account any acceleration due to the external torque on the bump. If we assume that the crossing time is not much shorter than $t_{\rm cross} = 280\Omegainv$, then our proposed SI criterion would be met --- according to \citet{Li_2021} the timescale of filaments is around $\sim 200\Omegainv$. \textit{However}, the SI might not outpace the GI. In \texttt{low-res,A=0.6}, the Roche density is reached at around the $t = 200\Omegainv$ anyway. Considering that there is some delay between the beginning of the simulation and reaching the $Z/\Pi$ needed for the SI, there does not seem to be enough time for the SI to act before the GI forms planetesimals.

\subsection{Are these pressure bumps Rossby-wave unstable?}
\label{sec:discussion:RWI}

Indeed, they very well might be. In \citet{Carrera_2021} we calculated that the largest Rossby-wave stable pressure bump is probably not much larger than $A \approx 0.20$. The ones in this paper are significantly larger than that. But this only serves to make the situation more precarious for the SI and mm-size grains. We pushed the bump past the RWI limit, right to the edge of the GI, and still the SI could not produce strong clumping for mm grains. A more realistic (i.e., smaller) pressure bump would have concentrated particles less and would have retained stronger headwind $\Pi$.

\subsection{Should we run \texttt{high-res,A=0.5} longer?}
\label{sec:discussion:longer-run}

It is possible that if we were to prolong \texttt{high-res}, strong clumping might appear later. Indeed, \citet{Yang_2017} showed that increasing resolution and increasing the runtime led to strong clumping for lower $Z/\Pi$ than was seen in \citet{Carrera_2015}. Comparing our total runtime of $t = 265\Omegainv$ to past works we find that at that point filaments were clearly visible in \citet{Li_2021} but barely visible for \citet{Yang_2017}. Considering that our particle size is slightly larger, our resolution is higher, our $Z$ is much larger, and our $\Pi$ is much smaller, we feel that evidence for strong clumping should have appeared by now.

Importantly, there is a strong argument \textit{against} prolonging the simulation: It might create artefacts. More specifically, this simulation has already run long enough to allow particles to cross the entire box. Running a simulation for much longer than the box crossing time would allow particles to wrap around the simulation box multiple times. This might lead to artefacts and unphysical results. For example, allowing particles to wrap around the box multiple times would imply that particles go through a chain of large pressure bumps that cover most of the disk. While indeed observations show the presence of multiple bumps, as the particles drift inward toward regions of higher gas density, the Stokes number should decrease. This is not captured in our local setup.

Clearly, the next step is not to prolong \texttt{high-res} but to move the simulation out of the shearing frame and into a global disk model.

%%%%%%%%%%%%%%%%%%%%%%%%%%%%%%%%%%%%%%%%%%%%%%%%%%
%
%   CONCLUSIONS
%
%%%%%%%%%%%%%%%%%%%%%%%%%%%%%%%%%%%%%%%%%%%%%%%%%%
\section{Summary and Conclusions}
\label{sec:conclusions}

In this paper we have tested the limits of planetesimal formation. Our \texttt{high-res,A=0.5} simulation is the highest-resolution 3D simulation of the SI performed to date. In it, we inserted a population of mm-size dust grains and the largest axisymmetric pressure bump that is just shy of the size that would lead to a gravitational instability in the dust. Our key result is simple:

\begin{itemize}
\item The $Z/\Pi$ we measured is well above the critical limit where previous studies \citep{Carrera_2015,Yang_2017,Li_2021} found clumping would occur if this value held over an extended radial region of the disk.

\item We did not observe any strong clumping nor planetesimal formation.
\end{itemize}

The likely reason is that the region with high-$Z/\Pi$ (i.e., the region conductive to the SI) is narrow and the particle crossing time $t_{\rm cross}$ across this region is shorter than the growth timescale of the SI $t_{\rm grow}$. Whatever progress the particles make toward strong clumping is lost when they leave the high-$Z$ particle pileup induced by the bump. This points to an additional criterion for planetesimals to form by the SI: $t_{\rm cross} > t_{\rm grow}$ --- i.e., the residence time of particles within a high-$Z/\Pi$ region must be long enough to allow the SI to develop filaments.

This result significantly limits the possible pathways for planet formation:

\begin{itemize}
\item Either protoplanetary disks routinely form grains larger than 1~mm,

\item The popular model in which planetesimals form when an axisymmetric pressure bump triggers the SI is incorrect.
\end{itemize}

For example, one possible alternative is that planetesimals form in vortices \citep[e.g.,][]{Barge_1995,Lyra_2015,Raettig_2021}. Either way, our results call for taking a hard look at how we understand planet formation, and at the computer models we use to study it. Our results also highlight the importance of future instruments, such as the ngVLA, that can determine whether cm-size grains are abundant at large distances from the star. \citep{2020tnss.book...25M}

%------------------------------%
% ACKNOWLEDGEMENTS
%------------------------------%
\begin{acknowledgements}
We thank the anonymous referee for their constructive comments and new insights. DC acknowledges Anders Johansen and Michiel Lanbrechts for a very fruitful discussion about the role of residency time and interpretation of our results. DC and JBS acknowledge support from NASA under {\em Emerging Worlds} through grant 80NSSC21K0037. The numerical simulations and analyses were performed on {\sc Stampede 2} through XSEDE grant TG-AST120062.
\end{acknowledgements}

%%%%%%%%%%%%%%%%%%%%%%%%%%%%%%%%%%%%%%%%%%%%%%%%%%
%
%   BIBLIOGRAPHY
%
%%%%%%%%%%%%%%%%%%%%%%%%%%%%%%%%%%%%%%%%%%%%%%%%%%

%\bibliography{sample631}{}
\bibliography{references}{}
\bibliographystyle{aasjournal}

\end{document}